# 3D-Hopkinson bar : new experiments for dynamic testing on soils

J.F Semblat
*Laboratoire Central des Ponts et Chaussées, Paris, France*

M.P. Luong, G. Gary
*Laboratoire de Mécanique des Solides, Ecole Polytechnique, Palaiseau, France*

ABSTRACT : The direct analysis of the dynamic response of materials is possible using Split Hopkinson pressure bar method. For soils, it has to be adapted since the specimen has generally poor mechanical properties. An original experimental arrangement called *"Three-Dimensional Split Hopkinson Pressure Bar" (3D-SHPB)* is proposed. It allows the measurement of the complete three-dimensional dynamic response of soils. Different types of confinement systems are used. The results on different loading paths are compared with other works on sand and clay. The analysis at grain-size level gives further elements on the comminution process.

## 1. FAST LOADINGS ON SOILS

### 1.1 Introduction

In the field of soil dynamics, many different methods and problems are considered. However, there is no real unified approach to investigate similarly as various problems as : earthquake engineering, pile driving, dynamic compaction, vibratory isolation... These diverse problems involve various frequency ranges or strain magnitudes. An attempt of classification is proposed in figure (1) comparing different practical problems and tests in terms of frequency, strain magnitude and ratio between wave length and dimensions of the domain (or the specimen). For small values of this ratio $\lambda/l_{ref}$, wave propagation phenomena prevail [13] (see figure (1)). Otherwise, they may be neglected and the « dynamic » behaviour of the material can be directly analysed.

Both approaches are presented in [13] : direct analysis of dynamic response of soils [13,15] and study of wave propagation phenomena in soils [13,16]. In this paper, we focus on experimental studies dealing with dynamic soil response : first experiments in the 60's without real control of the transient loading, Hopkinson bar based methods in the 70's and 80's and our 3D-Split Hopkinson Pressure Bar (3D-SHPB) [13,15].

### 1.2 First dynamic experiments

W.Heierli [9] performed dynamic experiments (falling mass) in an oedometric device considering one-dimensional assumption. However, he has not taken into account wave propagation phenomena in the experimental device itself. Comparisons between static and dynamic cases show very different responses for low densities and close results for dense specimens. W.Heierli also tried to make a link with pulse propagation experiments in loose soils.

R.V.Whitman [20] investigated dynamic shear loadings on sand. From Terzaghi works, he considered that, for fast shearing of sand, *grains cannot choose the mean resistance path* whereas they can find it for slow shear. Furthermore, he thought the quantitative evaluation of various response parameters is not satisfactory.

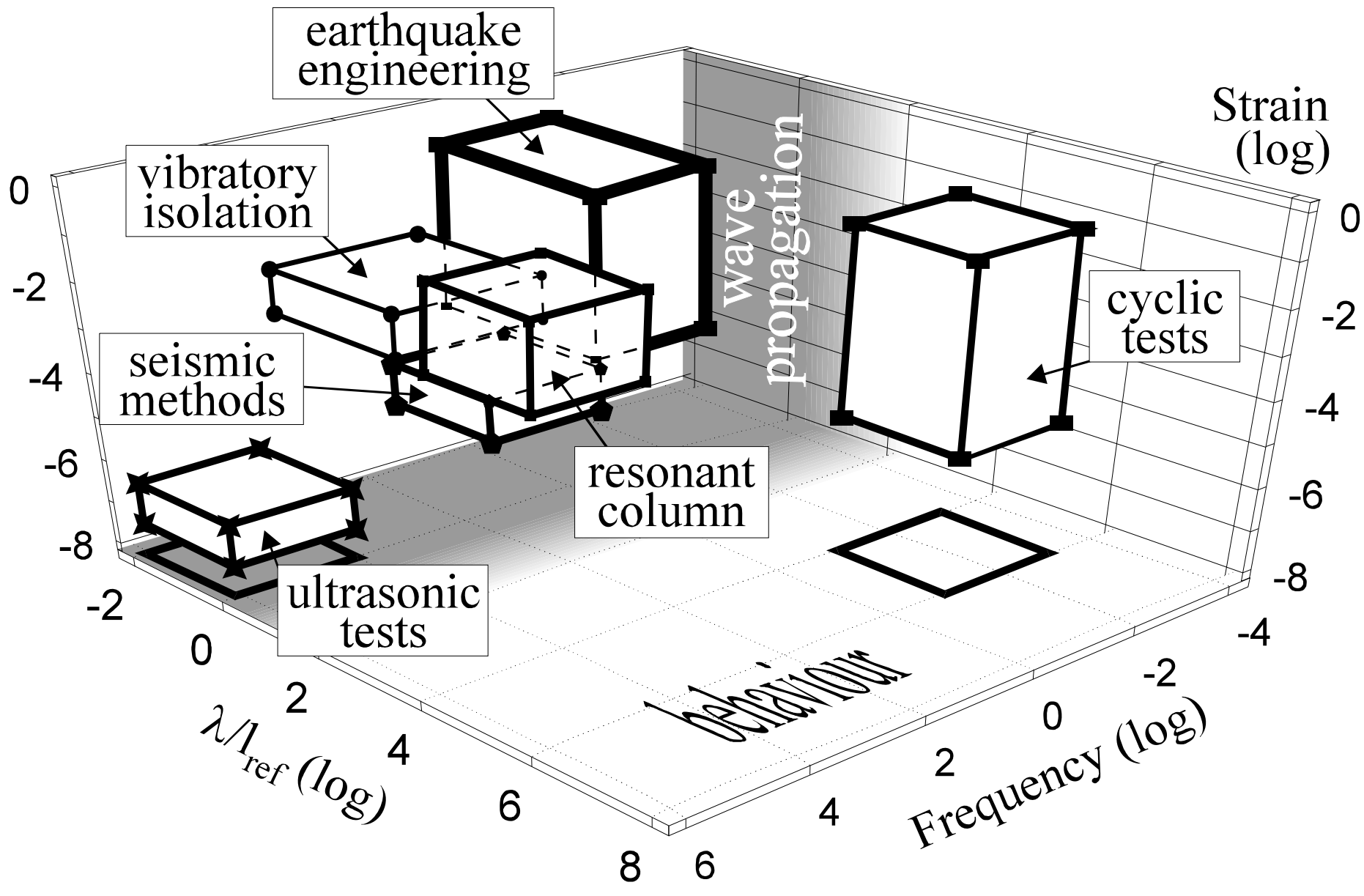

Figure 1 : *Classification of different types of dynamic tests on soils*

### 1.3 Hopkinson type loadings

In the 70’s in France, G.Aussedat and J.Meunier [10] developped dynamic tests on soils (falling mass with fast filming of the crushing phase). They also performed experiments based on Split Hopkinson Pressure Bar method and well-adapted to soil testing. They designed a low impedance bar (nylon) because their first experiments on steel bars were very disappointing.

Using clay specimen, these two experimental arrangements (dynamic crushing and Hopkinson bars) lead them to the following results :

- **crushing experiments** : J.Meunier performed the analysis with fast filming of the tests. This technique allows the study of *plastic wave propagation* in the specimen during crushing
- **Hopkinson bars testing** : experiments were made on very thin (1 to 10 mm) clay specimens, friction was important because the diameter of the bars is 36 mm. The only transmitted wave was taken into account in these tests, it did not allow the determination of the stress in the specimen (stress in the transmitter bar is less than 2 MPa). G.Aussedat and J.Meunier performed these experiments with and without confining pressure (from 0.2 to 0.6 MPa). From their experiments, amplitude of the transmitted wave was increasing for higher impact speeds or decreasing specimen thicknesses. However, there was *no influence of the confining pressure* on transmitted wave amplitude. It seems to be logical considering specimen thicknesses and values of confining pressure.

Their work is one of the first to investigate dynamic response of soils with an acurate control of transient phenomena in the experimental device itself. Split Hopkinson Pressure Bar tests (SHPB) seem to be well-adapted for soils and allows a good control of wave propagation phenomena.

## 2. THE "CLASSICAL" S.H.P.B METHOD

### 2.1 Experimental arrangement

The original Hopkinson device (with only one cylindrical bar) was modified by Kolsky (two bars) for indirect measurements on both sides of the specimen. The "classical" Split Hopkinson Pressure Bar system is then composed of two axial bars (incident bar and transmitter bar) and a striker bar launched by a gas gun. Figure (2) gives a schematic of this "classical" SHPB device.

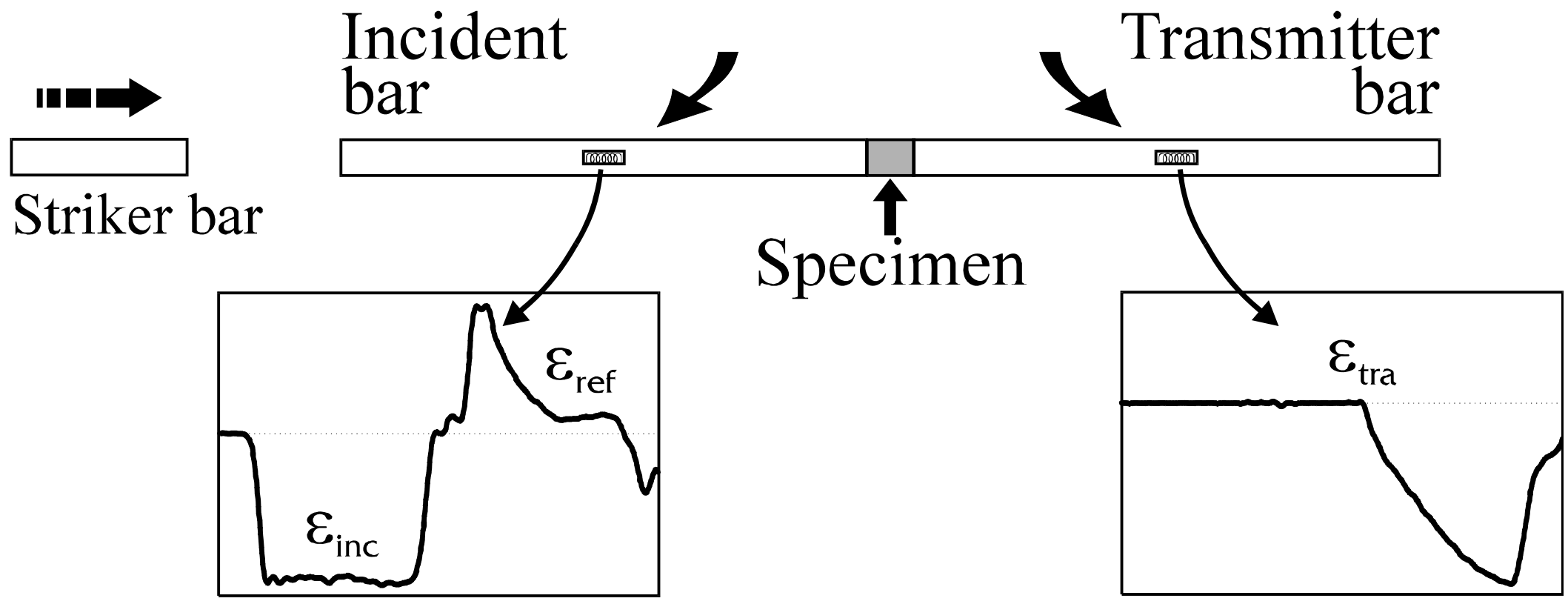

Figure 2 : *"Classical" Split Hopkinson Pressure Bar device.*

As shown in figure (2), the specimen is put between the two main bars. The impact between the striker bar and the incident bar generates a compressive stress wave (loading wave and unloading waves). The main characteristic of Hopkinson type experiments is to perform indirect strain measurements : strains are measured on the bars (and not directly on the specimen). Gauges give the values of incident ($\varepsilon_{inc}$), reflected ($\varepsilon_{ref}$) and transmitted ($\varepsilon_{tra}$) strain waves in the bars (see fig. (2)). From these measurements, it is possible to determine in every point of the bars and at every time the values of forces and displacements (stress and strain). It is especially the case for bar-specimen interfaces.

## 2.2 Dynamic loading

### *2.2.1 Axial stress in the specimen*

Propagation of the stress wave in the bars and at both bar-specimen interfaces is an important aspect of dynamic experiments on S.H.P.B device. On both bar-specimen interfaces, a process of multiple reflections and transmissions takes place. It depends on the mechanical parameters of both bar and specimen. A 3D-schematic is given in [13,14,15] depicting the variations of axial stress with time and location. It indicates clearly that axial stress in the specimen increases progressively. This phase of the experiment is called the "transient phase" during which propagation phenomena strongly prevail. Afterwards axial stress becomes more and more uniform along the specimen. This is the main interest of SHPB method : it allows *uniform stress distribution under high strain rates*.

### *2.2.2 Two main experimental phases*

Dynamic tests on Split Hopkinson Pressure Bar can generally be divided in two main stages :

- a "transient phase" : the first reflections and transmissions of the loading wave lead to a non homogeneous stress state. Wave propagation phenomena in the specimen strongly prevail so that this stage of the test is called the "transient phase". Incident force is much more higher than transmitted force (see fig. (3))
- a "fast quasi-static phase" : after several reflections and transmissions of the loading wave on both interfaces, a stress equilibrium state along the specimen length is reached. This stage of the test is called the "fast quasi-static" phase : the axial stress is homogeneous in the whole specimen (on Fig. (3), incident and transmitted forces are balanced).

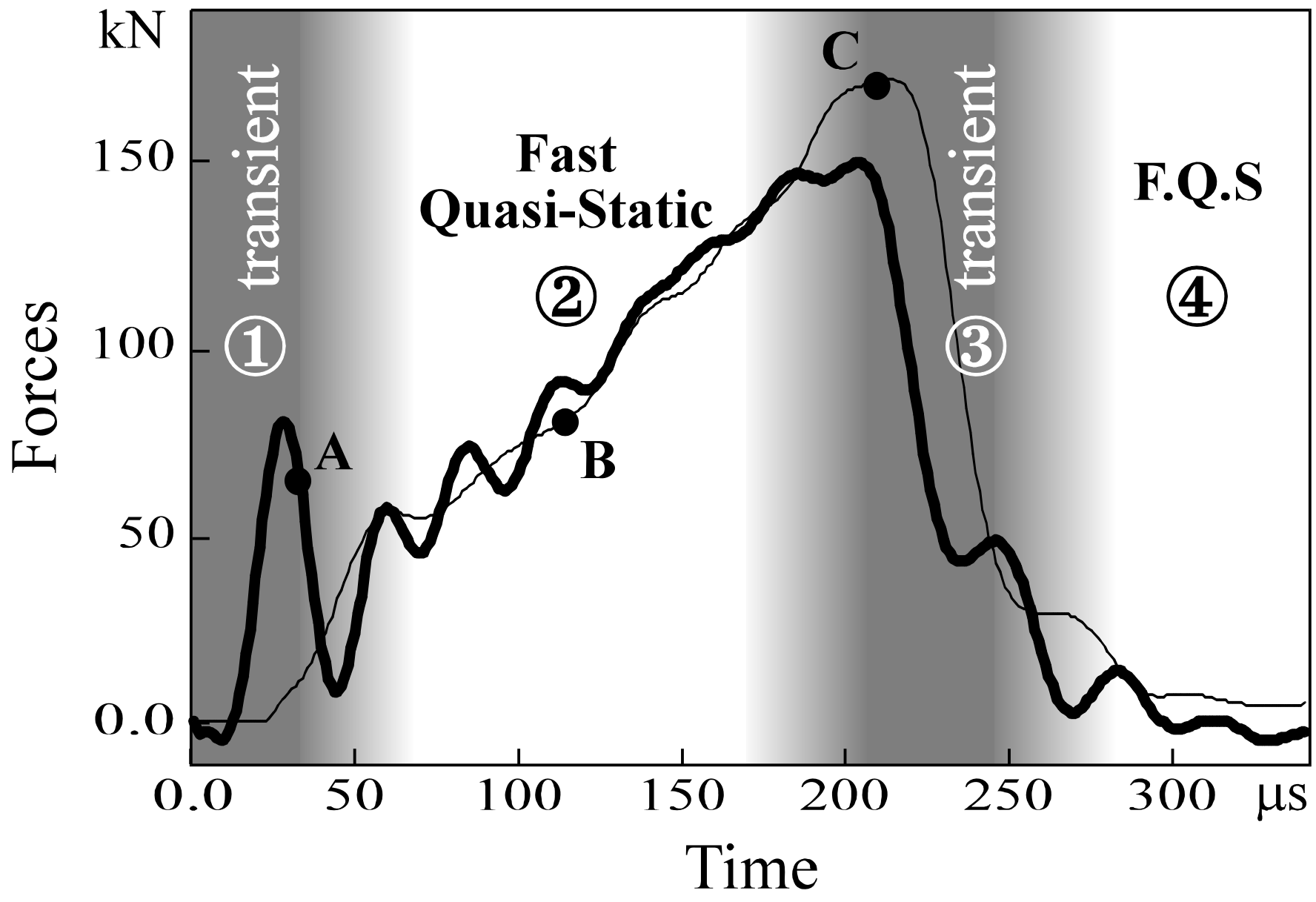

Figure 3 : *Forces at both specimen faces showing two loading phases (resp. unloading).*

The curves in figure (3), giving forces versus time, can be depicted considering particular points :

- point A : incident force is positive and transmitted force is zero, there is a loading force on the upper face of the specimen whereas the loading wave has not reached the lower face yet. There is *no equilibrium of loading forces* in the specimen
- point B : incident and transmitted forces are equal, the *specimen is in equilibrium* . The axial stress calculated with these force values is very close to real axial stress in the whole specimen length
- point C : incident force is lower than transmitted force, incident force starts to decrease while unloading wave has not reached the lower face yet. There is *no equilibrium of unloading forces* in the specimen

From figure (3), it is obvious that both incident and transmitted forces are equal after the initial transient phase. These experimental results given by dynamic experiments on soils show that the classical assumption of S.H.P.B method is encountered : it is possible to perform high strain rate experiments on soils since there is a "Fast Quasi-Static" phase allowing direct determination of the dynamic response (behaviour) of this soil. However, recent analysis techniques give much more information about transient phases [11].

### 2.3 Determination of mechanical parameters

#### *2.3.1 Fictitious wave carrying*

As it is shown in figure (2), strain waves are measured on the bars : incident and reflected waves on the incident bar and transmitted wave on the transmitted bar. However, to determine forces and displacements at both bar-specimen interfaces, strain waves have to be fictitiously carried to the interfaces. The most important is to identify the starting point of each strain wave. Zhao [21] gives many explanations on this point and the methods to perform an acurate determination of these points. Elastic simulation of strain wave propagation in the specimen allows for example a more precise identification (dispersive phenomena in the bars being also taken into account).

#### *2.3.2 Strain and stress in the specimen*

In the bars, behaviour and propagation parameters are readily related. This is the main advantage of the S.H.P.B method. Since dispersive phenomena are corrected [7], the assumption of one-dimensional propagation is fully justified.

The expressions of axial stress and strain are then given as follows :

$$\sigma_{ax}=\rho.C_0.v \quad \text{and} \quad \varepsilon_{ax}=\frac{v}{C_0} \qquad (1)$$

where ρ is the density, $C_0$ the wave velocity and *v* the particle velocity.

These expressions are valid for every type of propagation medium. For purely elastic bars, expression (1) takes the following form :

$$\sigma_{ax}=\rho.C_0^2.\varepsilon_{ax} \qquad (2)$$

As strains are measured on the bars, forces at both bar-specimen interfaces are deduced from measured strain waves :

$$F_{inc}(t)=E.S_b.[\varepsilon_{inc}(t)+\varepsilon_{ref}(t)]$$
$$F_{tra}(t)=E.S_b.\varepsilon_{tra}(t)$$

Axial stress $\sigma_{ax}$ in the specimen is then derived from these expressions :

$$\sigma_{ax}=\frac{S_bE}{2S_{spec}}\left[\varepsilon_{inc}(t)+\varepsilon_{ref}(t)+\varepsilon_{tra}(t)\right] \qquad (3)$$

where $S_b$ is the section of the bars and $S_{spec}$ the section of the specimen.

The expression (3) is only valid for the "fast quasi-static" phase of the test. It is calculated from the forces at both faces of the specimen : this is a good assumption if both forces are equivalent (specimen equilibrium).

### 2.4 Recent dynamic experiments

Many recent researches provide results from Hopkinson type experiments. C.W.Felice and G.E.Veyera in USA [2,3,4,19], S.Shibusawa in Japan [17] and A.M.Bragov in Russia [1] performed SHPB tests to determine soil response under high strain rate. As soil constitutive parameters are generally poor, the classical experimental device has to be adapted to this particular material.

C.W.Felice [2,3,4] performs Hopkinson bar experiments on saturated sand (or clayey sand) specimens. These are oedometric dynamic tests (Felice used a confining cylinder) without measurements of radial stress. The main goal of this work is to improve the analysis of experimental results in the initial loading phase. For soil specimen, homogeneisation delay of the stress can be important and the determination of response parameters at the beginning of loading can be difficult. Slenderness of the specimens is less than 0.2 for a bar diameter of Φ=60.3 mm (see table (I)). Experimental results indicate that specimens *resistance increases because of saturation*. Veyera and Ross [19] works concern sand specimen under undrained dynamic compression using a thick-walled container. Specimens length ranges from l=6.3 to l=12.7 mm and the bar diameter is Φ=50.8mm. The strain rates involved in these experiments are from 1000 to 2000 $s^{-1}$ (see table (I)).

In Japan, Shibusawa and Oida [17] investigate dynamic response of soils (mainly clays) to study the influence of water content and specimen dimensions. The experimental device allows measurement of incident and reflected waves only. The transmitted force is measured directly on the specimen (see table (I)). Shibusawa and Oida give prominence to an exponential increase of the dynamic modulus with increasing water content.

A.M.Bragov [1] studies dynamic response of plasticine in jacket-confined tests. The bars diameter is Φ=20 mm and the specimen length is l=15 mm. This is, with our work, the first research to investigate three-dimensional dynamic response by performing circumferential strain measurement. Bragov uses four strain gauges on the jacket to determine radial strain of the specimen during dynamic axial loading (see table (I)).

Main characteristics of the tests performed by these different authors are collected in table (I). Experimental devices are all Hopkinson type systems (with one only bar for Shibusawa [17]). The dimensions of bars, specimens and the confining methods used are varied : confining cylinder, confining pressure (see table (I)). The results of our approach will be compared with those obtained by the authors listed in table (I).

| Authors | type of soil | specimen dimensions | striker & bars | loading duration | type of confinement |
|---|---|---|---|---|---|
| Meunier (1974) | clay | l=2 à 15mm<br>Φ=36mm | $l_{stri}$=0.15m<br>nylon<br>Φ=36mm | 500µs | air pressure |
| Felice (1985,91) | clayey sand and alluv. | l=13/25mm<br>Φ=60.9mm | $l_{stri}$=0.25m<br>metal<br>Φ=60.9mm | 125µs | thick cylinder |
| Bragov (1994) | plasticine (⇔ clay) | l=15mm<br>Φ=20mm | $l_{stri}$=0.25m<br>metal<br>Φ=20mm | 200µs | thin cylinder<br>th.=10mm<br>measur. of $\varepsilon_r$ |
| Veyera (1995) | dry or saturated sand | l=6.3/12.7mm<br>Φ=50.8mm | $l_{stri}$=0.65m<br>metal<br>Φ=50.8mm | 257µs | thick cylinder :<br>th.=25mm |
| Shibusawa (1992) | silt+clay+ sand | l=50/100mm<br>Φ=50mm | $l_{stri}$=0.25m<br>metal<br>Φ=25mm | 80µs | none |
| Semblat, Luong, Gary (1995) | dry sand | l=10,15,20,25 & 30mm<br>Φ=40mm | $l_{stri}$=0.85m & 0.5m<br>metal & PMMA<br>Φ=40mm | 350µs | thick cylinder<br>air pressure<br>oil pressure |

Table I : *Different Hopkinson type dynamic tests on soils*

## 3. "3D-SHPB" : A NEW DEVICE FOR DYNAMIC TESTING ON SOILS

### 3.1 Experimental apparatus

For the dynamic testing of soils, it is necessary to modify the classical Hopkinson arrangement : Meunier proposed a nylon bars device, Felice the use of a rigid confining cylinder, Bragov jacket-confined experiments (see table (I)). However, considering the influence of stress path on soils response, it would be very interesting to measure (or control) both axial and radial stresses. The dynamic response could then be analysed following the three-dimensional stress paths. In this study, oedometric dynamic tests using a rigid confining cylinder are carried out on a special device called *"Three-Dimensional Split Hopkinson Pressure Bar" (3D-SHPB).* When using a rigid confining cylinder, zero radial strain can be ensured while radial stress cannot be correctly estimated. Using a radial bar in contact with the specimen through the confining cylinder, this special device allows measurement of radial stress with time [13,14,15]. Figure (4) gives a schematic of the "Three-Dimensional Split Hopkinson Pressure Bar".

The special device showed in figure (4) involves three Hopkinson type bars :

- 2 axial bars to measure axial displacements and forces on both sides of the specimen (as for classical SHPB method)
- 1 radial bar to evaluate the radial stress during the test

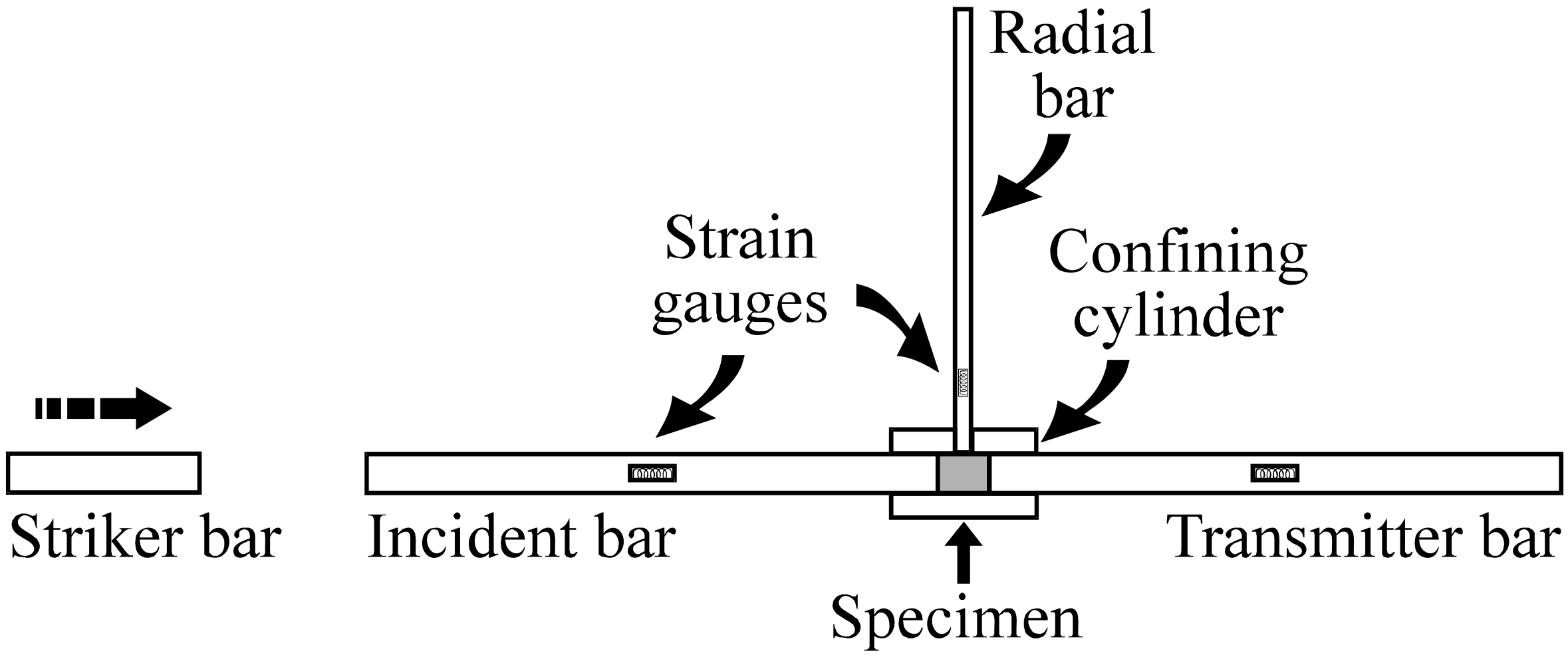

Figure 4 : *Three-Dimensional Split Hopkinson Pressure Bar device.*

Mechanical and geometrical characteristics of the axial bars are given in table (II) :

| **bar properties** | |
|---|---|
| diameter | $\Phi$= 40 mm |
| length | $l_b$= 2 m |
| Young modulus | $E$= 70000 MPa |
| density | $\rho_b$= 2820 kg/m$^3$ |
| velocity | $C_0$=5180 m/s |
| **striker properties** | |
| diameter | $\Phi$= 40 mm |
| length | $l_s$= 0.85 m |
| Young modulus | $E$= 70000 MPa |
| mass | $m$= 3.012 kg |
| impact speed | *2 to 20 m/s* |

Table II : *bar and striker bar properties*

## 3.2 Rigid confinement tests (dynamic oedometric)

### *3.2.1 Axial dynamic response*

All the specimens are composed of dry Fontainebleau sand. Density of the specimens is constant : $\rho$=1667 kg/m$^3$. The tests performed on the experimental device shown in figure (4) are called "rigid confinement tests" : the rigid confining cylinder prevents from radial strain. The confining cylinder must therefore be sufficiently rigid or thick to give a small radial strain. This is verified from radial stress measurements and numerical results given in [14].

| test | spec. length | impact speed | strain rate | modulus | mean modul. |
|---|---|---|---|---|---|
| sadur001 | **10 mm** | 3.4 m.s$^{-1}$ | 393 s$^{-1}$ | 468 MPa | 476 MPa |
| sadur002 | | | 473 s$^{-1}$ | 479 MPa | |
| sadur003 | | | 497 s$^{-1}$ | 482 MPa | |
| sadur006 | | 5.8 m.s$^{-1}$ | 771 s$^{-1}$ | 377 MPa | 426 MPa |
| sadur007 | | | 725 s$^{-1}$ | 443 MPa | |
| sadur008 | | | 697 s$^{-1}$ | 457 MPa | |
| sadur011 | | 9.9 m.s$^{-1}$ | 1245 s$^{-1}$ | 440 MPa | 460 MPa |
| sadur012 | | | 1190 s$^{-1}$ | 476 MPa | |
| sadur013 | | | 1188 s$^{-1}$ | 464 MPa | |
| sadur016 | **15 mm** | 3.4 m.s$^{-1}$ | 345 s$^{-1}$ | 515 MPa | 509 MPa |
| sadur017 | | | 314 s$^{-1}$ | 502 MPa | |
| sadur018 | | | 279 s$^{-1}$ | 511 MPa | |
| sadur021 | | 5.8 m.s$^{-1}$ | 468 s$^{-1}$ | 582 MPa | 591 MPa |
| sadur022 | | | 458 s$^{-1}$ | 602 MPa | |
| sadur023 | | | 446 s$^{-1}$ | 588 MPa | |
| sadur026 | | 9.9 m.s$^{-1}$ | 793 s$^{-1}$ | 593 MPa | 589 MPa |
| sadur027 | | | 821 s$^{-1}$ | 604 MPa | |
| sadur028 | | | 827 s$^{-1}$ | 571 MPa | |
| sadur031 | **20 mm** | 3.4 m.s$^{-1}$ | 220 s$^{-1}$ | 648 MPa | 700 MPa |
| sadur032 | | | 240 s$^{-1}$ | 719 MPa | |
| sadur033 | | | 268 s$^{-1}$ | 732 MPa | |
| sadur036 | | 5.8 m.s$^{-1}$ | 379 s$^{-1}$ | 557 MPa | 570 MPa |
| sadur037 | | | 361 s$^{-1}$ | 582 MPa | |
| sadur038 | | | 359 s$^{-1}$ | 570 MPa | |
| sadur041 | | 9.9 m.s$^{-1}$ | 634 s$^{-1}$ | 616 MPa | 626 MPa |
| sadur042 | | | 640 s$^{-1}$ | 619 MPa | |
| sadur043 | | | 602 s$^{-1}$ | 644 MPa | |
| sadur046 | **25 mm** | 3.4 m.s$^{-1}$ | 200 s$^{-1}$ | 704 MPa | 693 MPa |
| sadur047 | | | 222 s$^{-1}$ | 679 MPa | |
| sadur048 | | | 223 s$^{-1}$ | 695 MPa | |
| sadur049 | | 5.8 m.s$^{-1}$ | 318 s$^{-1}$ | 655 MPa | 661 MPa |
| sadur050 | | | 316 s$^{-1}$ | 649 MPa | |
| sadur051 | | | 317 s$^{-1}$ | 679 MPa | |
| sadur052 | | 9.9 m.s$^{-1}$ | 508 s$^{-1}$ | 657 MPa | 661 MPa |
| sadur053 | | | 521 s$^{-1}$ | 664 MPa | |
| sadur054 | | | 536 s$^{-1}$ | 661 MPa | |
| sadur055 | **30 mm** | 3.4 m.s$^{-1}$ | 167 s$^{-1}$ | 686 MPa | 697 MPa |
| sadur056 | | | 148 s$^{-1}$ | 670 MPa | |
| sadur057 | | | 158 s$^{-1}$ | 735 MPa | |
| sadur058 | | 5.8 m.s$^{-1}$ | 264 s$^{-1}$ | 684 MPa | 700 MPa |
| sadur059 | | | 270 s$^{-1}$ | 684 MPa | |
| sadur060 | | | 247 s$^{-1}$ | 733 MPa | |
| sadur061 | | 9.9 m.s$^{-1}$ | 444 s$^{-1}$ | 635 MPa | 657 MPa |
| sadur062 | | | 430 s$^{-1}$ | 692 MPa | |
| sadur063 | | | 455 s$^{-1}$ | 645 MPa | |

Table III : *Experimental results for all oedometric dynamic tests*

Figure (5) plots axial stresses versus axial strains for three different rigid confinement tests (oedometric). These three tests correspond to a same specimen length (l=10mm) but to different strain rates (see figure (5) and table (III)). For rigid confinement tests, axial responses are nearly linear for loading and unloading phases with different slopes in both cases but equivalent ones from one test to another. The elastic part of the response is not really clear as the global dynamic behaviour of sand is shown to be highly anelastic (figure (5)). The oedometric response is compared further with responses on other kind of loading paths.

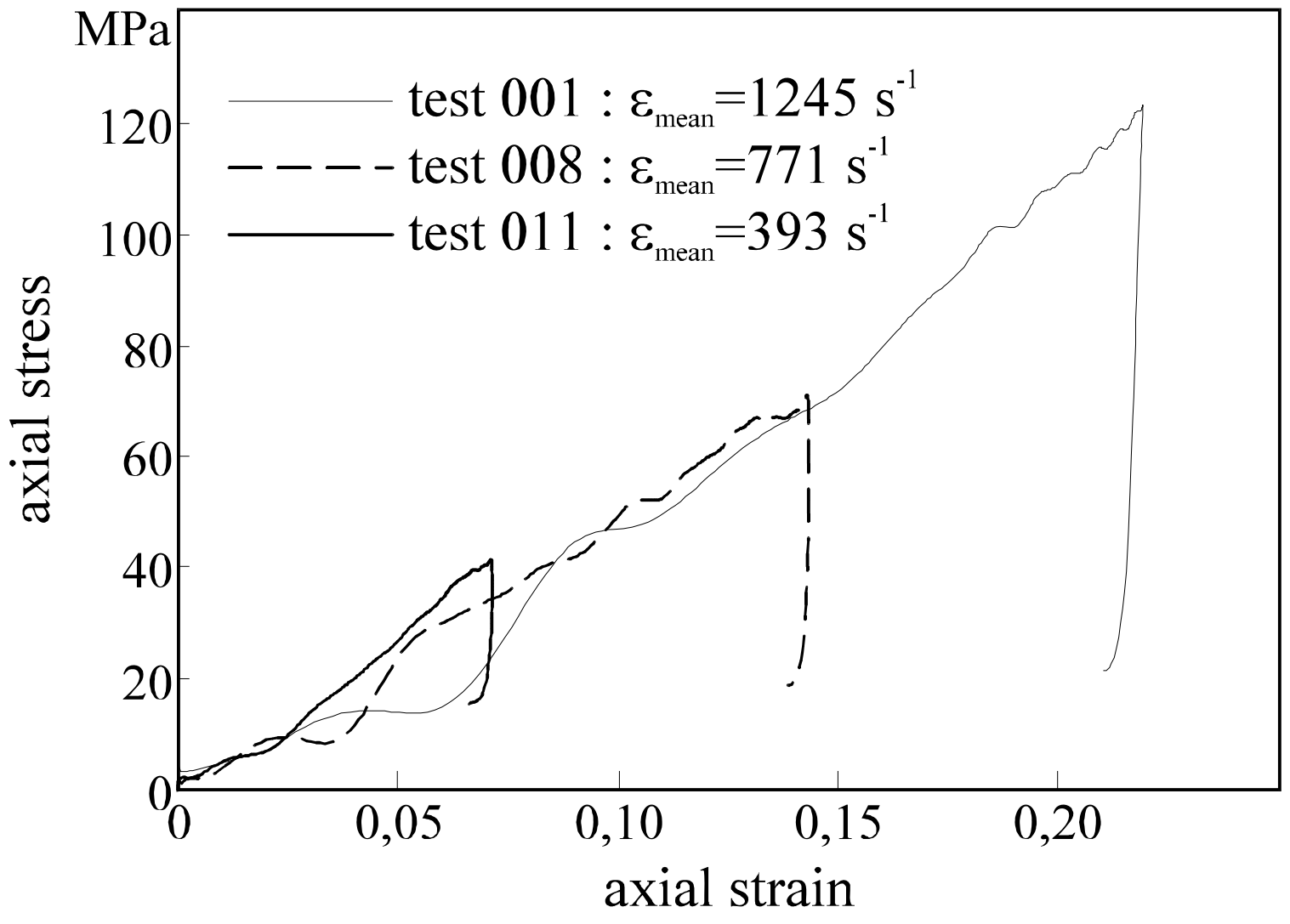

Figure 5 : *axial dynamic response on 3D-Split Hopkinson Pressure Bar*

*3.2.2 Dynamic moduli of sand*

The oedometric dynamic tests are performed using specimen of five different lengths and three impact speeds (of the striker bar on the incident bar), each test being repeated three times identically. For all these tests, values of corresponding strain rates and dynamic moduli are given in table (III). Strain rate values range from 200 $s^{-1}$ up to 1245 $s^{-1}$ depending on the impact speed (of the striker bar on the incident bar) and on the specimen length. Slopes of the dynamic stress-strain curves refer to a highly anelastic, but linear, dynamic response and range from 350 MPa to 750 MPa approximately (see table (III)). Lowest values (350-450 MPa) correspond to highest strain rates (800-1200 $s^{-1}$) and highest values (650-750 MPa) to lowest strain rates (200-500 $s^{-1}$).

From the experimental results, it is clear that variations of these slopes are not negligeable at all, but the relationship with strain rate values is not obvious. The analysis of three-dimensional aspects of the dynamic response gives interesting results concerning the potential dynamic effect.

*3.2.3 Radial stress measurement*

The confining pressure is not constant during axial dynamic loading. To quantify the variations of radial stress with time, an original experimental arrangement is proposed. 3D-SHPB device, presented in figure (4), allows the measurement of the stress wave in the radial bar from which the radial stress in the specimen is derived (see expressions (1) and (3)). It is then possible to compare values of axial and radial stress with time. Mechanical and geometrical properties of the radial bar are given in table (IV).

Because of the stiffness of the confining cylinder, there is a strong variation of radial stress during axial loading. As it is shown in figure (6), radial and axial stresses increase simultaneously during the major part of the loading phase. Afterwards, the radial stress starts to decrease whereas axial stress still increases (see figure (6)). For the unloading phase, radial and axial stresses are both decreasing very fast.

| radial bar | |
|---|---|
| diameter | *Φ= 40 mm* |
| length | $l_b$= *1.442 m* |
| Young modulus | *E= 94000 MPa* |
| density | $\rho_b$= *8520 kg/m*$^3$ |
| velocity | $C_0$=*3323 m/s* |

Table IV : *radial bar properties*

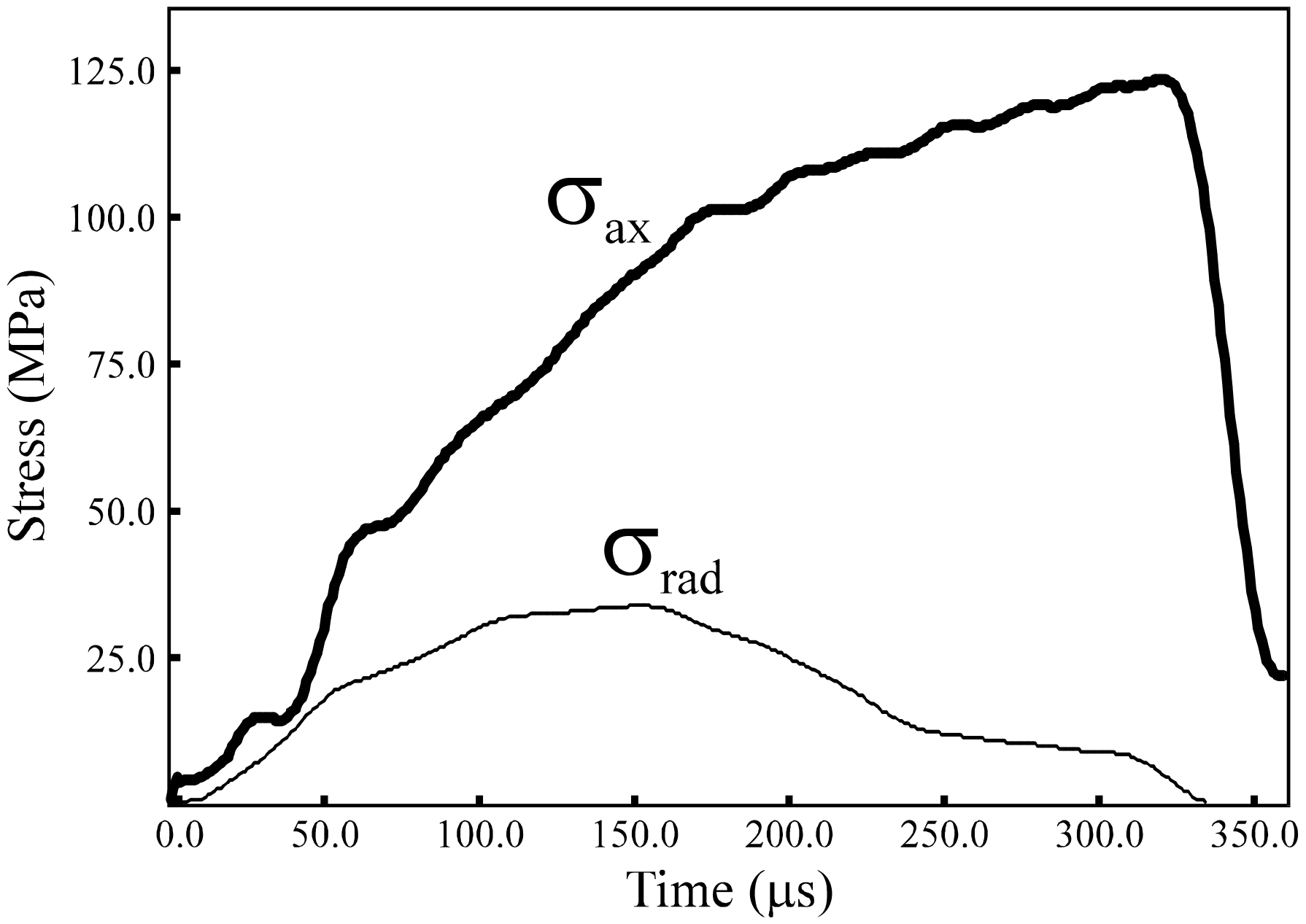

Figure 6 : *Variations of axial and radial stress with time.*

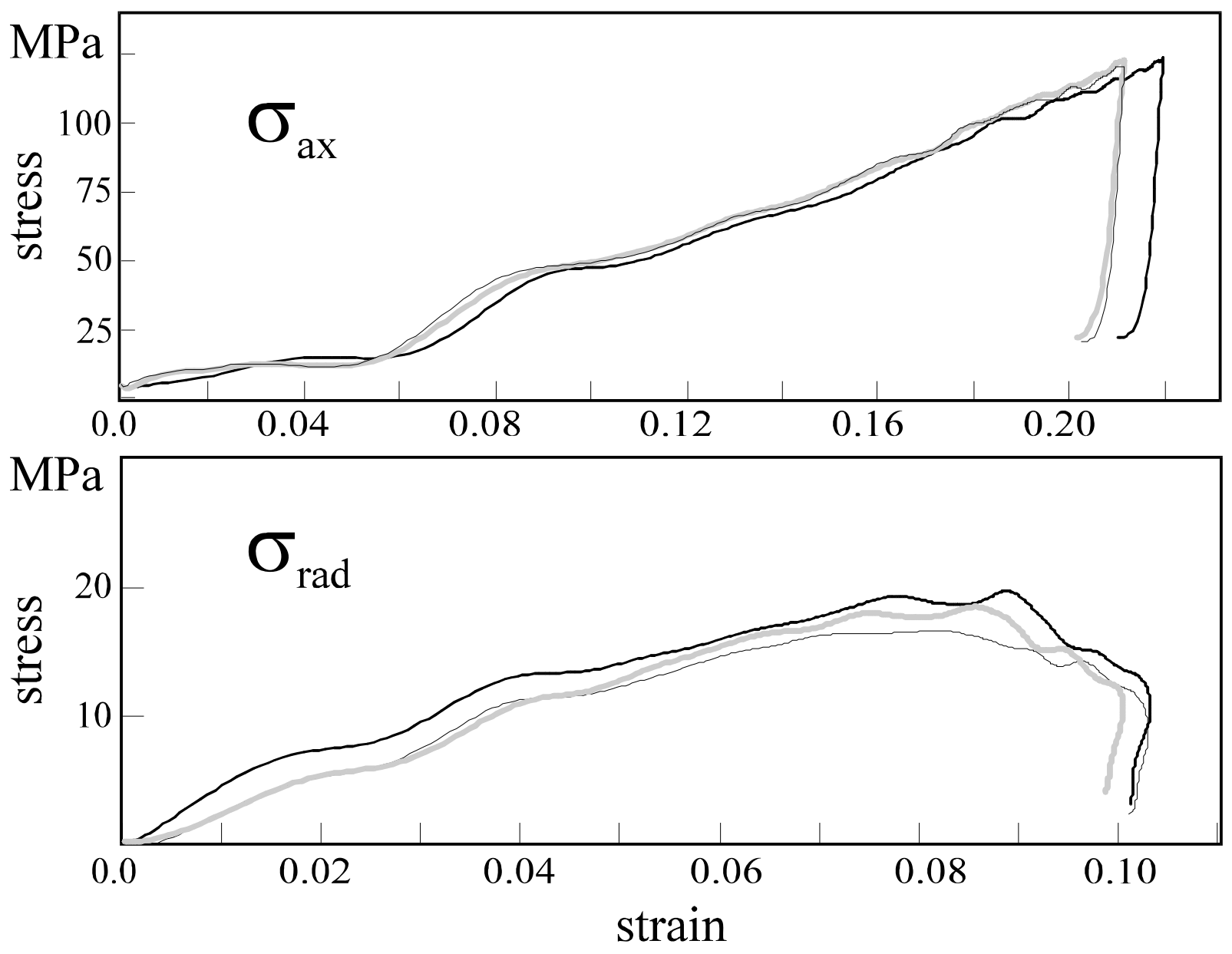

Figure 7 : *Axial and radial stresses versus strain for three identical tests*

The variations of confining pressure (radial stress) for oedometric dynamic tests must be considered. They are very important during axial loading : radial stress $\sigma_{rad}$ is variable with time as, for this test, $\sigma_{rad}$ reaches a maximum value of about 30 MPa for a time t=150 μs (see figure (6)).

*3.2.4 Test reproducibility*

Experimental reproducibility is studied by repeating each tests (specimen length, strain rate) three times identically. It is very good for axial stress and acceptable for radial stress measurements (see figure (7)). Experimental results given in table (III) are also good towards this point (values of strain rates for identical tests). It is then possible to study the complete loading paths for rigid confinement tests (oedometric dynamic tests).

### 3.3 Three-dimensional aspects of dynamic loading

*3.3.1 Mean and deviatoric stresses*

Starting from the axial and radial stress measurements, it is possible to evaluate the three-dimensional loading path in terms of mean stress "p" and deviatoric stress "q". The analysis of such mechanical parameters is of much interest for soils. Calculation of mean and deviatoric stresses leads to the following expressions :

$$p = \frac{\sigma_{ax} + 2\sigma_{rad}}{3} \tag{4}$$

$$q = \sigma_{ax} - \sigma_{rad} \tag{5}$$

where $\sigma_{ax}$ and $\sigma_{rad}$ are the axial and radial stresses respectively.

Figure (8) curves reveal clearly that, for a linear strain path ($\varepsilon_q/\varepsilon_v$=2/3 in oedometric tests), the stress path is also linear. However, loading slope and unloading slope on p-q diagrams are different. The structure of the specimen is actually different after the loading phase (grain crushing, see section (6)).

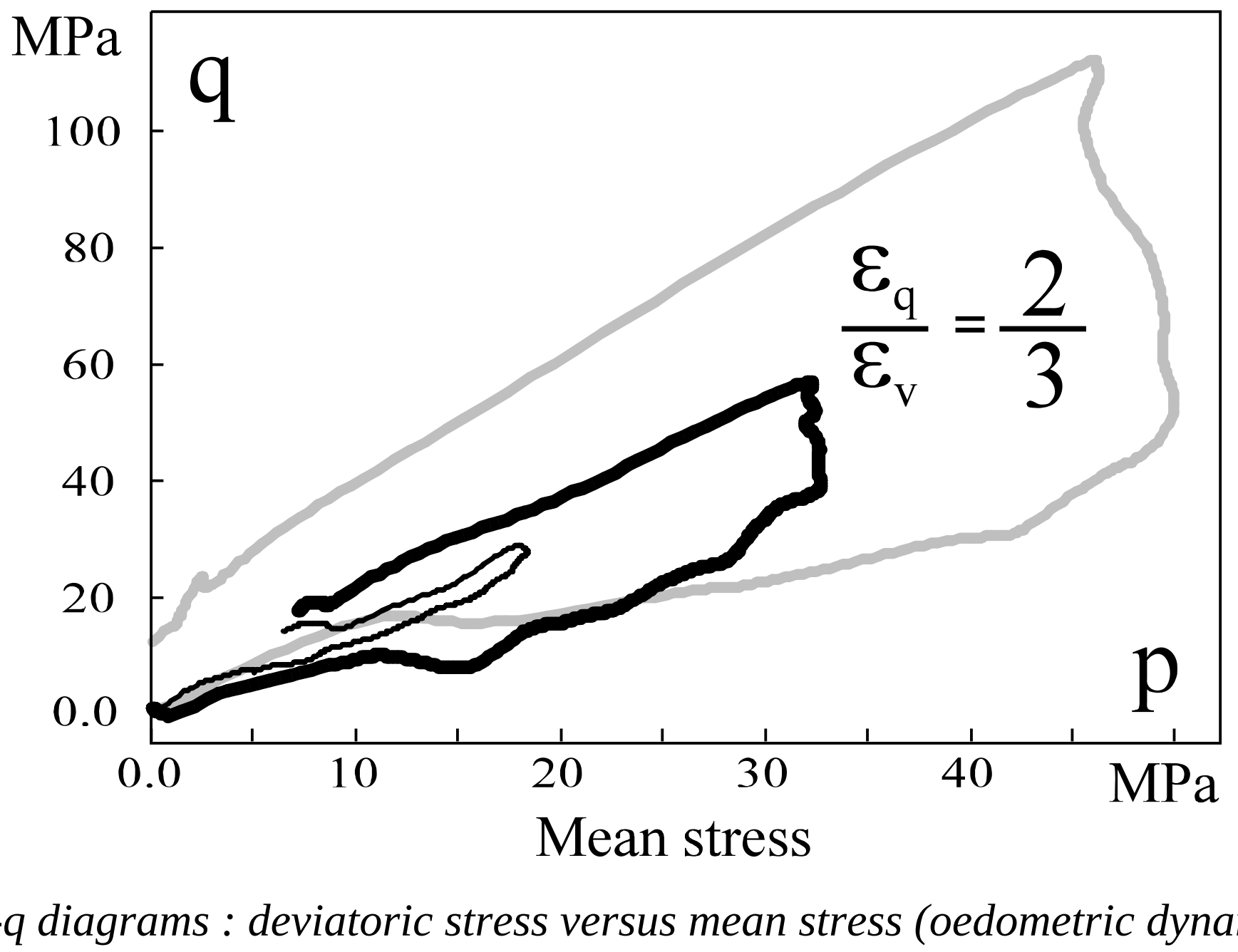

Figure 8 : *p-q diagrams : deviatoric stress versus mean stress (oedometric dynamic tests)*

The influence of stress paths on the material response is well known for static loading but was never clearly analysed in case of dynamic experiments. Comparisons of axial and radial stresses give more quantitative results in the next paragraph. In the following section, other tests are performed under constant and slightly variable confining pressures. It allows the comparison of the stress paths for various confining conditions under high strain rates.

*3.3.2 Pseudo Poisson's ratio*

From experimental measurements of axial and radial stresses, it is possible to investigate further the three-dimensional dynamic response. Considering the oedometric strain paths used and the quasi-linear aspect of the axial dynamic response, a *dynamic pseudo Poisson's ratio* can be calculated using theory of elasticity. Stress and strain tensors are then very simply related :

$$\varepsilon_{22} = \frac{\sigma_{22}}{E} - \frac{\nu}{E}.(\sigma_{11} + \sigma_{33}) \quad \text{for example}$$

For rigid confinement tests, we assume that principal stress and strain directions are the same than axial and radial bars directions. As these tests are oedometric, $\varepsilon_{22}$ is zero, $\sigma_{11}=\sigma_{rad}$ and $\sigma_{22}=\sigma_{33}=\sigma_{ax}$ (where $\sigma_{ax}$ and $\sigma_{rad}$ are the axial and radial stresses respectively). The previous relationship is simplified as follows :

$$\sigma_{rad} - \nu.(\sigma_{ax} + \sigma_{rad}) = 0$$

It leads to the expression of the *dynamic pseudo Poisson's ratio*

$$\nu = \frac{\sigma_{rad}}{\sigma_{ax} + \sigma_{rad}} \tag{6}$$

From the experimental measurements of axial and radial stresses and the previous expression of pseudo Poisson's ratio, corresponding numerical values are derived.

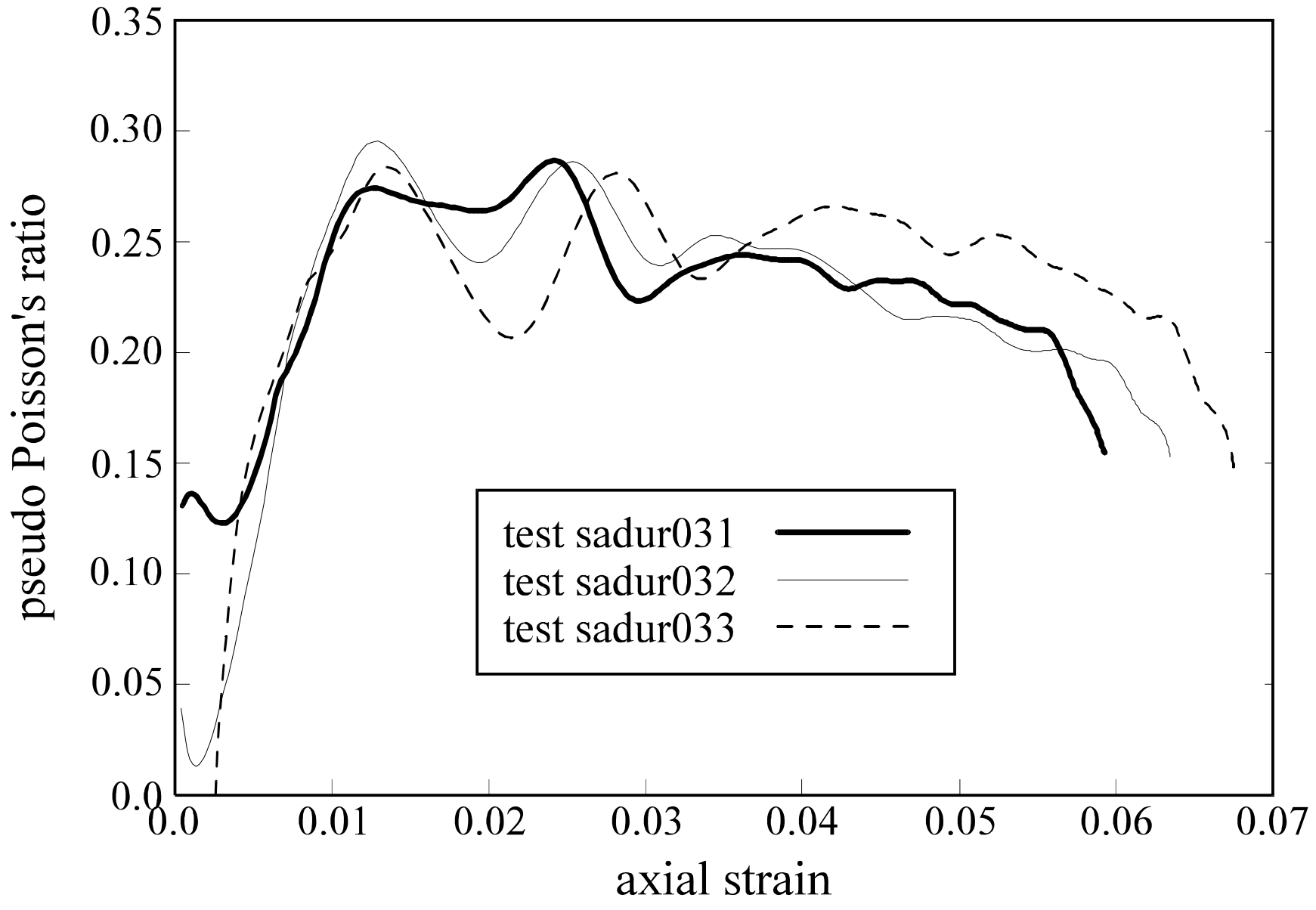

Figure 9 : *Dynamic pseudo Poisson's ratio for oedometric strain paths*

Using expression (6), experimental measures of axial and radial stresses allow the calculation of the dynamic pseudo Poisson's ratio $\nu_{dyn}$. Curves of figure (9) give values of $\nu_{dyn}$ for three identical tests (see table (III)). After the initial transient phase ($\varepsilon<0.015$), values of $\nu_{dyn}$ are nearly constant with axial strain. This is a very interesting result concerning this pseudo Poisson's ratio.

Values of $\nu_{dyn}$ given in table (V) correspond to various specimen lengths and strain rates. It appears from this values that the pseudo Poisson's ratio is the highest for highest values of strain rate (see table (III)). Dynamic radial confining effect is increasing with increasing strain rate.

| Test | Pseudo Poisson's ratio | Mean value |
|---|---|---|
| sadur011 | 0.327 | 0.367 |
| sadur012 | 0.387 | |
| sadur013 | 0.388 | |
| sadur021 | 0.322 | 0.308 |
| sadur022 | 0.304 | |
| sadur023 | 0.297 | |
| sadur031 | 0.244 | 0.243 |
| sadur032 | 0.239 | |
| sadur033 | 0.247 | |

Table V : *Dynamic pseudo Poisson's ratio for oedometric dynamic tests.*

## 4. OTHER LOADING PATHS

### 4.1 Various types of confinement

It is of much interest to compare the dynamic response of soils using different loading paths. In addition to the "rigid confinement" tests (oedometric tests on 3D-SHPB), three other types of confining systems are used [13] :

- *semi-rigid confinement* : the confining pressure applied to the specimen is not constant (uncompressible fluid).
- *soft confinement* : the soft confinement tests are performed with a compressible confining fluid ensuring a constant confining pressure during the tests.
- *low impedance tests* : all other tests are performed on duraluminium bars and the axial stress is high. For low impedance tests, the use of plexiglas bars allows low stress and low confining experiments. The mechanical impedance of that kind of bars is low : these experiment are called low impedance tests.

The confining cell presented in figure (10) gives a slightly variable (semi-rigid) or a constant (soft) confining for "semi-rigid" and "soft confining" tests. "Low impedance" tests are performed on a PMMA Hopkinson device and the specimen is confined with a constant pressure (same type of confining cell, see figure (10)). A special correction procedure allows to take into account damping and dispersive phenomena in the used viscoelastic bar (Zhao, 1992).

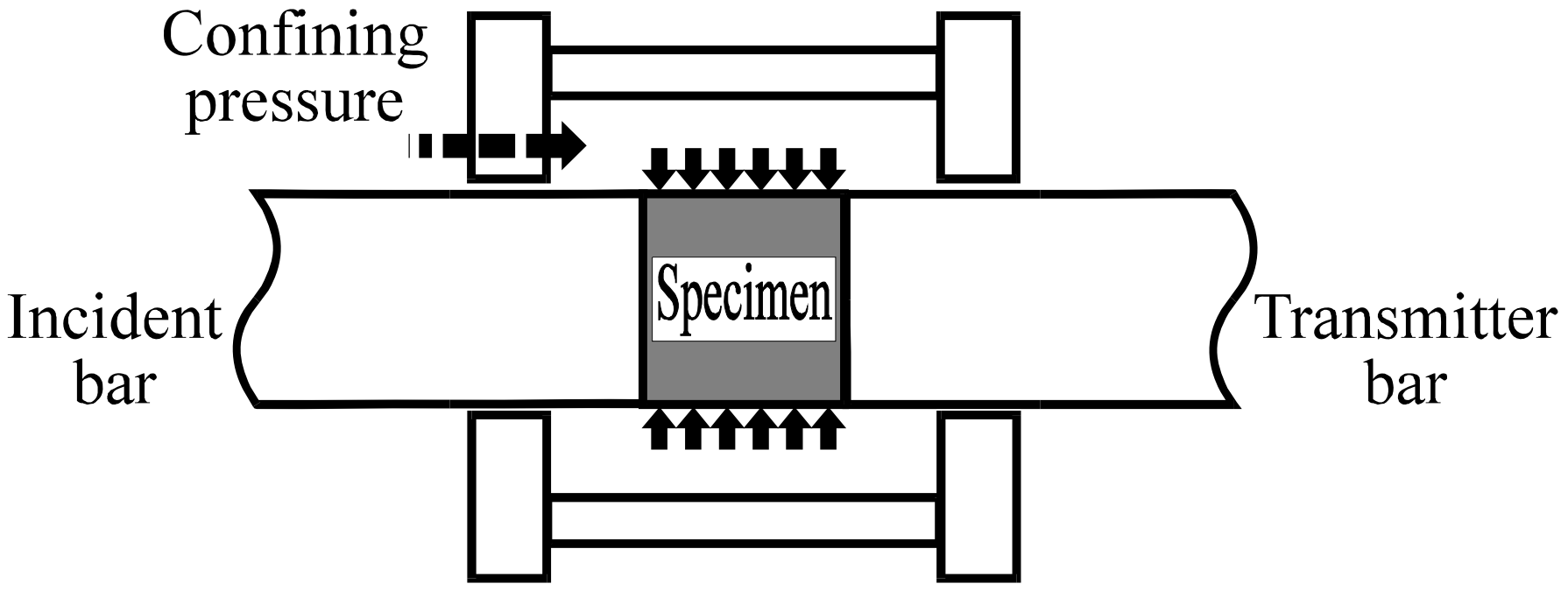

Figure 10 : *Confining cell for semi-rigid, soft confining tests and low impedance tests.*

### 4.2 Semi-rigid confinement tests

Semi-rigid confinement tests are performed under slightly variable confinement (uncompressible fluid) using the experimental arrangement presented in figure (10). The experimental device does not allow an acurate estimation of confining pressure variations (as for rigid confinement tests).

Figure (11) gives the axial stress versus axial strain for different semi-rigid confinement tests (slightly variable confinement). For that kind of tests, the dynamic response is nearly linear for both loading and unloading. Values of axial stress are of same order than for rigid confinement tests. Curves presented in figure (11) refer to tests performed under various confining pressure (3.0; 5.6 and 7.5 MPa) but with the same specimen length (l=10mm). Confining pressure has no strong influence on the dynamic response (for the present confining pressure values). It should be noted that this pressure may change during axial loading (no measurement of this changes is performed here).

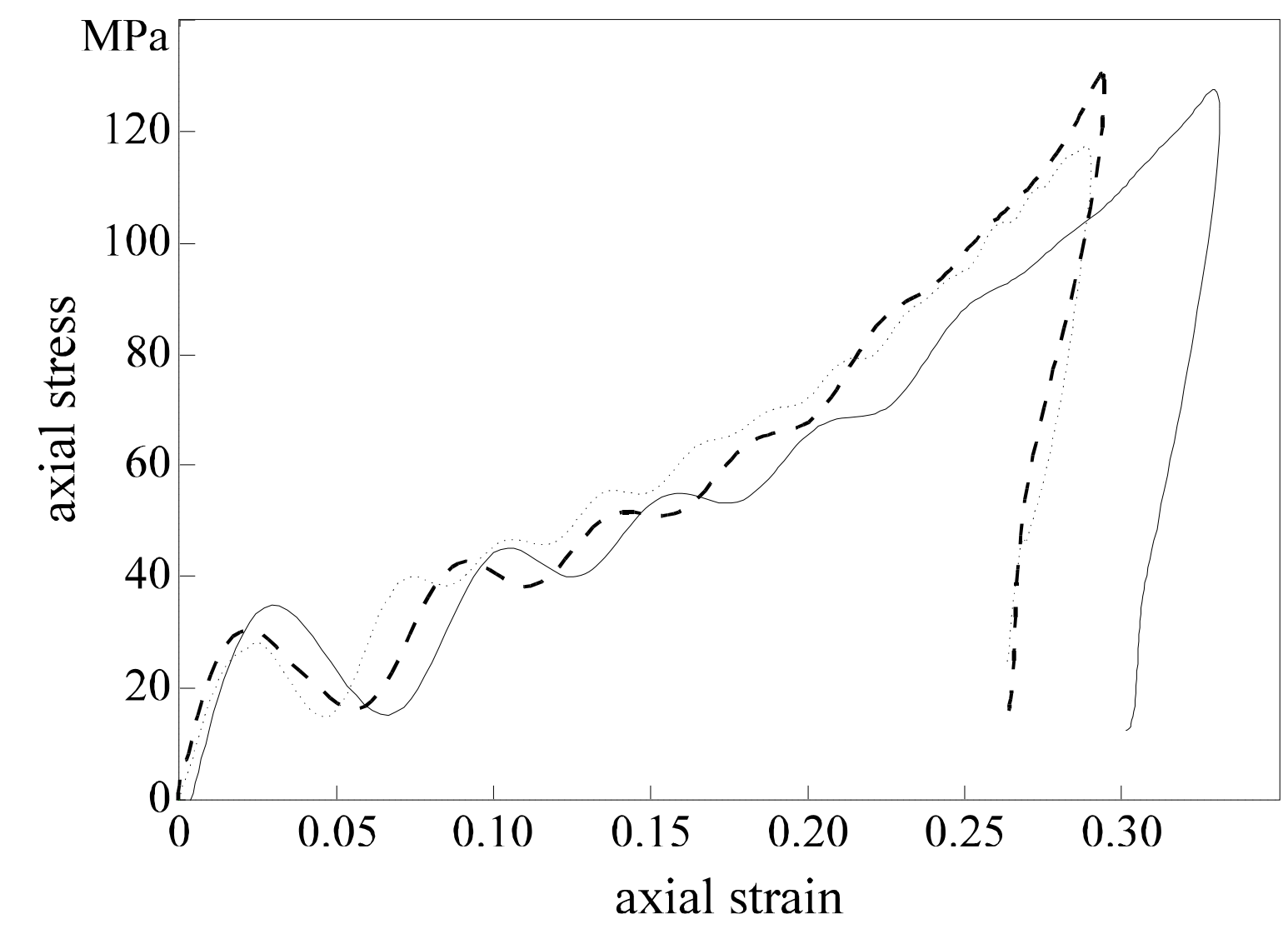

Figure 11 : *Axial stress versus axial strain (for semi-rigid confinement tests)*

### 4.3 Soft confinement tests

Soft confinement tests are performed under constant confining pressure (air pressure). It allows the comparison with the dynamic response of slightly variable confinement tests. Figure (12) gives the axial stress versus axial strain for a confining pressure of 2.5 MPa, the specimen length is l=11mm.

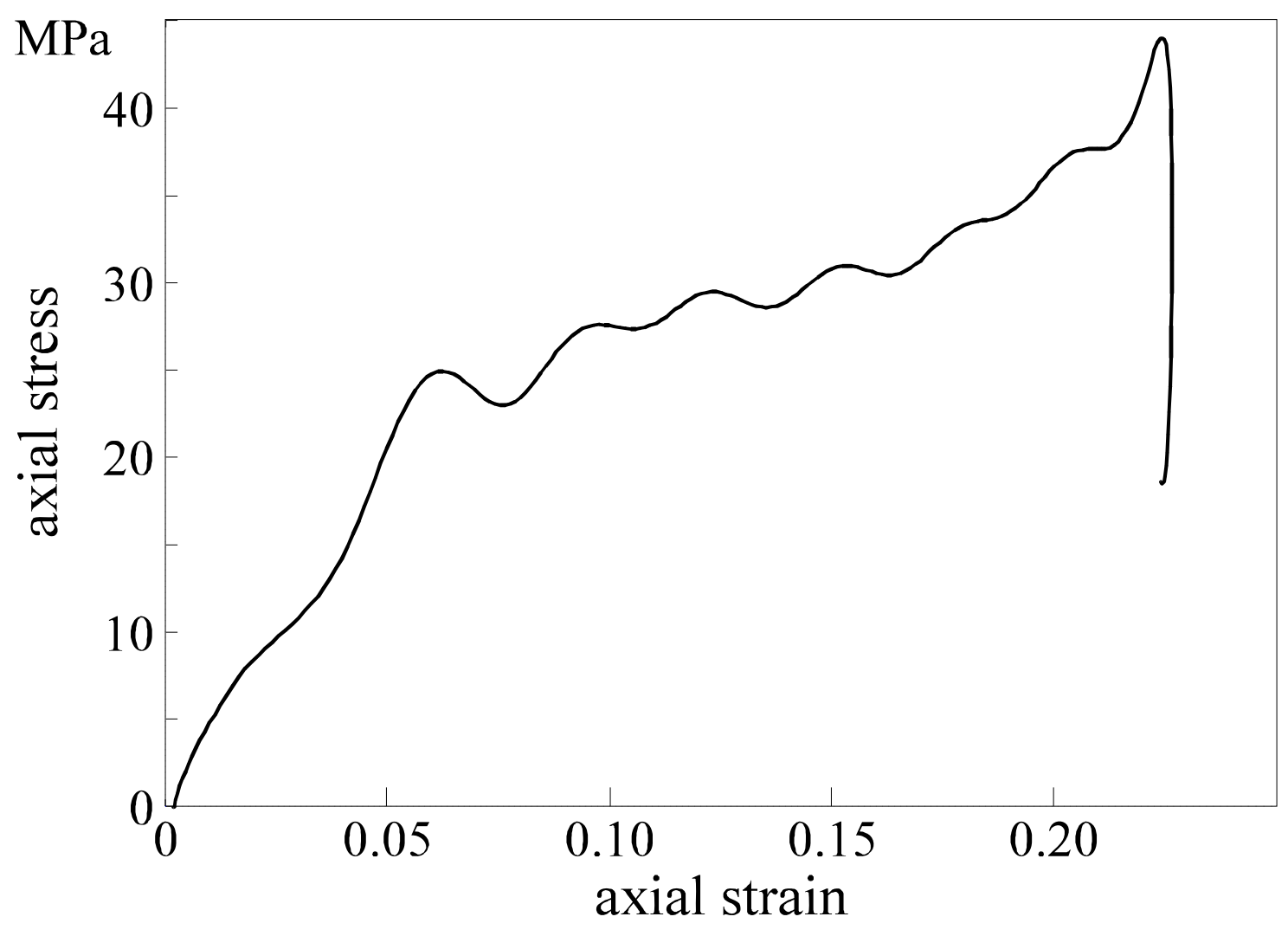

Figure 12 : *Axial stress versus axial strain (for soft confinement confinement tests)*

The maximum value of axial stress is 40 MPa, which is much lower than values of rigid or semi-rigid confinement tests. For soft confinement tests, the response is no more linear for the loading phase. For rigid and semi-rigid confinement tests, the variations of confinement during loading give a stiffened response of the material.

## 4.4 Low impedance tests (PMMA bars)

Low impedance tests are performed on PMMA bar (plexiglas). These bars have a much lower mechanical impedance than duraluminium bars used for all other tests (see table (VI)). Impedance ratio between the bars and the specimen is the lower. It allows a faster homogeneisation of the axial stress in the specimen.

| bar properties | |
|---|---|
| diameter | $\Phi$= 40 mm |
| length | $l_b$= 2 m |
| Young modulus | $E$= 6000 MPa |
| density | $\rho_b$= 1226 kg/m$^3$ |
| velocity | $C_0$=2210 m/s |

Table VI : bar properties for low impedance tests

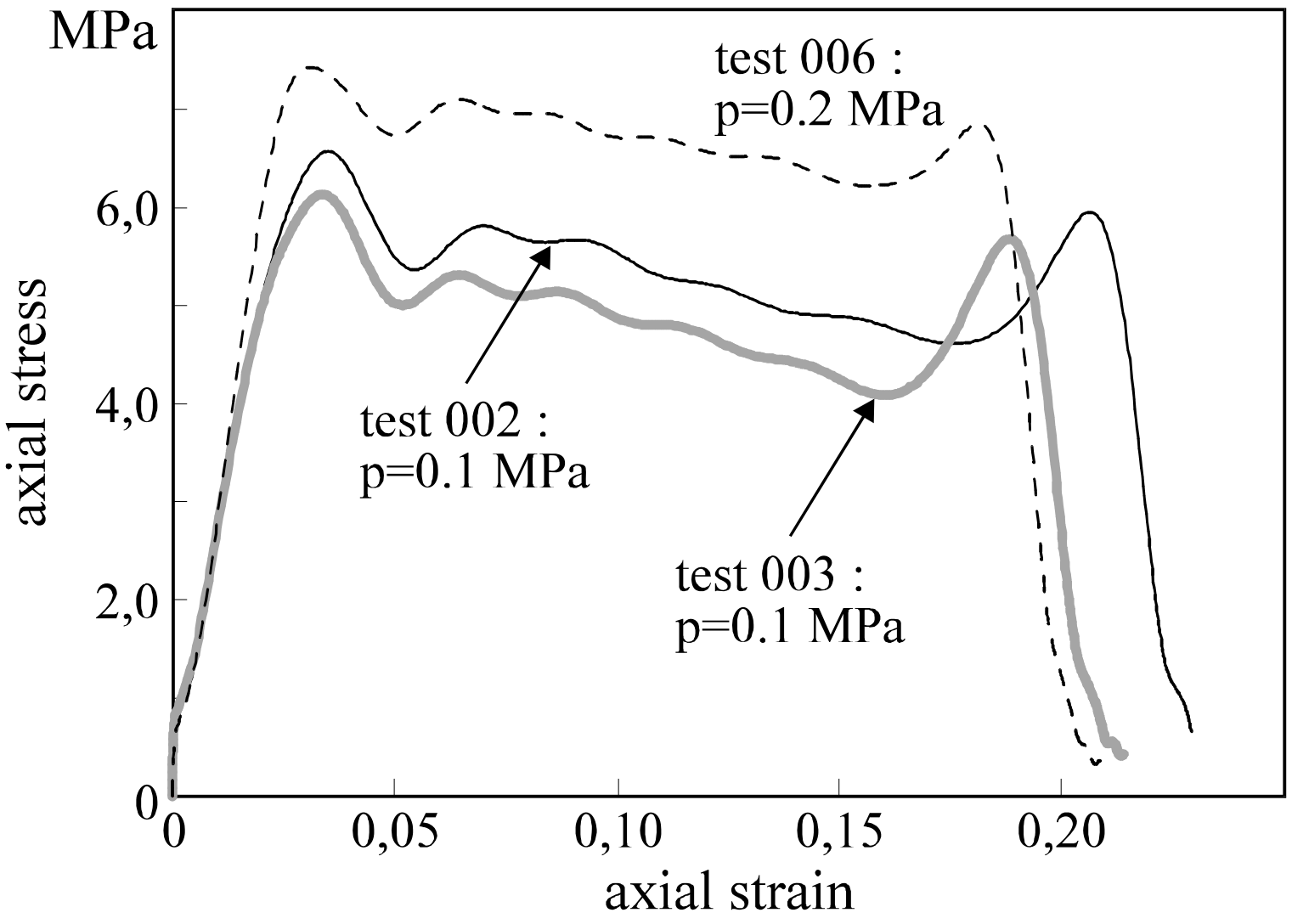

Figure 13 : *Axial stress versus axial strain (for low impedance tests)*

The experimental arrangement is the same than for soft and semi-rigid confinement tests. The strain measurements made on the bars are corrected by taking into account both geometrical dispersion in the bars and material dispersion due to plexiglas viscosity. This procedure is explained in details in the works of G.Gary and H.Zhao [7,21].

Figure (13) gives three curves axial stress versus axial strain from low impedance tests. The maximum axial stress is much lower than for other types of tests (less than 10 MPa). The dynamic response of the material is of softening type even if strain rates are of the same order as for the preceding tests.

## 5. COMPARISON OF THE DYNAMIC RESPONSES

### 5.1 Influence of the dynamic loading path

From the different dynamic tests performed, it is possible to appreciate the influence of stress path on the specimen response. The comparison of the dynamic responses on different loading paths (rigid confining, semi-rigid, soft and low impedance) is given in figure (14). From these curves, there is an obvious influence of the dynamic loading path on the dynamic response.

There are two kinds of dynamic response :

- for "low impedance" tests and "soft confining" tests : the specimen strength is decreasing during loading
- for "semi-rigid confining" tests and "oedometric tests" : behaviour is quite linear for loading. The increase of the confinement apparently strengthens the specimen under dynamic loading.

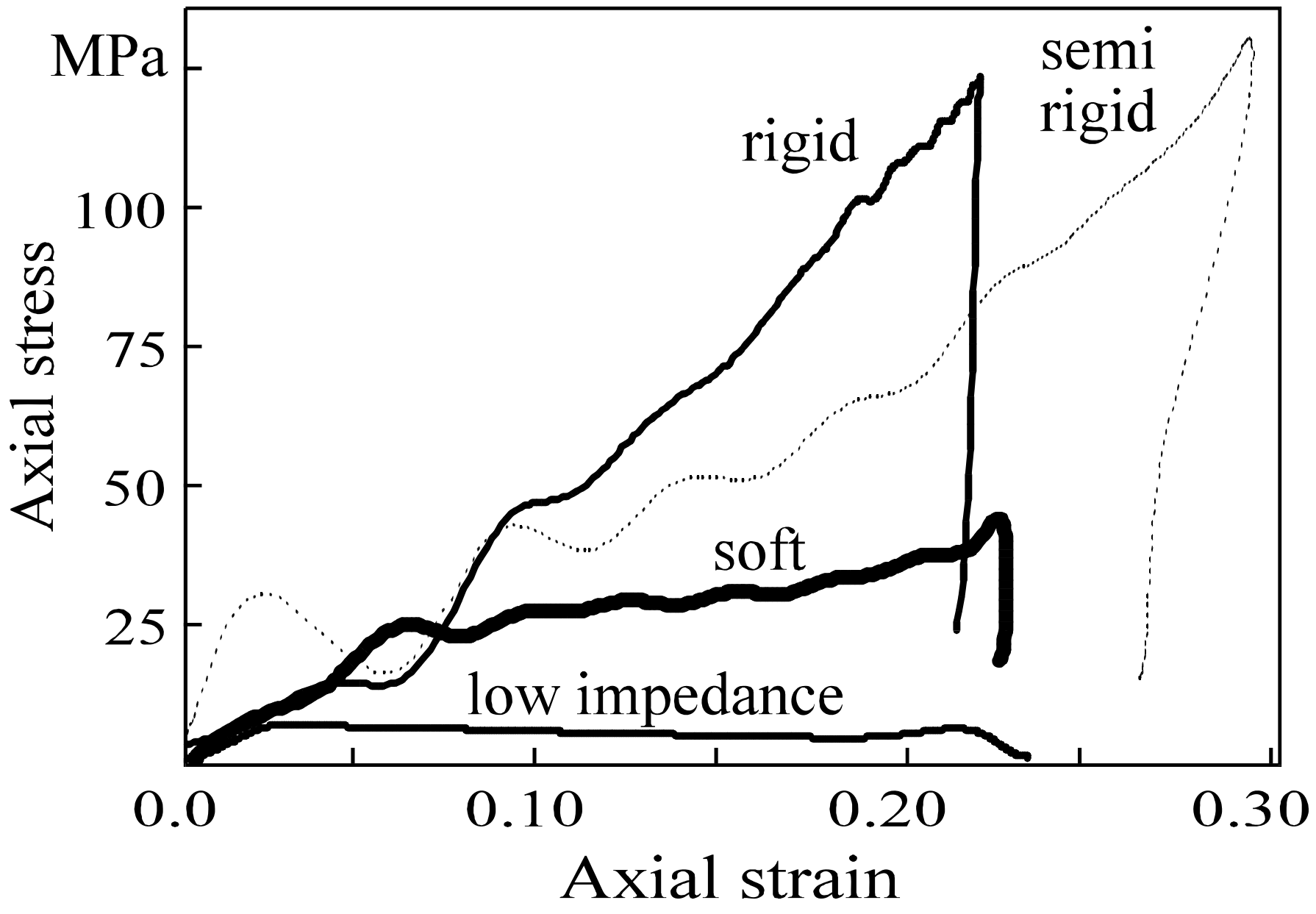

Figure 14 : *Axial stress vs axial strain corresponding to different types of confining conditions.*

### 5.2 Other experimental researches

As indicated at the beginning, other authors have studied soil response under fast loadings (see table (I)). It started with Heierli and Whitman in the late 50's, but there was no appropriate control of the dynamic processes in the experimental device itself.

Main results of more recent researches are collected in table (VI) : all authors use Hopkinson type experimental arrangements. Transient phenomena are well-controlled and a special confining procedure is generally used. The only work considering three-dimensional effects is due to Bragov but he studies different loading path than ours (his results concerns *dynamic strain paths*). Our experiments give dynamic soil responses on various stress paths using the same material (see table (VII).

The influence of saturation on the dynamic response is analysed by several authors. It seems to be important after the initial phase of closing of the voids (table (VII)). Our experiments do not investigate the influence of saturation but shows the effect of confinement stiffness on the dynamic response (soft, semi-rigid and rigid). Calculation of the dynamic pseudo Poisson's ratio $\nu_{dyn}$ is performed thanks to axial and radial stress measurements. The influence of strain rate on $\nu_{dyn}$ is shown. Analysis at grain-size level is also an original aspect of our research and it could be related in further works with experiments made by Shukla to study transient effects in granular forces variations (see §6).

| Authors | Main results and conclusions |
|---|---|
| Meunier (1974) | • first tests on nylon bars (low impedance) without correction of dispersive phenomena<br>• weak influence of confining pressure on the response |
| Felice (1985,91) | • bilinear behaviour:<br>* 1st phase for $\varepsilon <$ initial porosity, filling of the voids and crushing of the grains<br>* 2nd phase specimen fully saturated, response of the fluid |
| Bragov (1994) | • measurement of the circonferential strain from tests with soft confining cylinder on plasticine specimen |
| Veyera (1995) | • strong dependence of the dynamic response on the saturation index (stiffened of the behaviour with increasing saturation) |
| Shibusawa (1992) | • modulus increases with saturation<br>• one-bar test, results draught from incident and reflected waves only, axial stress possibly non homogeneous |
| Semblat, Luong, Gary (1995) | • Hopkinson bar experiments well-adapted for soils<br>• determination of the three-dimensional dynamic stress path, design of the *3D-S.H.P.B test*<br>• comparison of the dynamic responses on different loading paths (rigid, semi-rigid, and soft confinements)<br>• a variable confinement gives a higher stiffness of the dynamic response<br>• calculation of the dynamic pseudo Poisson's ratio<br>• analysis at grain-size scale : interesting results on the comminution process |

Table VII : *Main results for Hopkinson type dynamic tests given in table (I)*

## 6. ANALYSIS AT GRAIN-SIZE LEVEL

### 6.1 Grain-size changes

For all the "rigid confinement" tests, grain-size distributions of the specimens are compared : figure (15) gives a 3D diagram of the granulometric distributions versus maximal axial stress for all oedometric tests. This diagram clearly indicates that the percentages of large grain decrease whereas the percentages of small grain increase. Furthermore this qualitative remark, it is possible to quantify the variations of particle size.

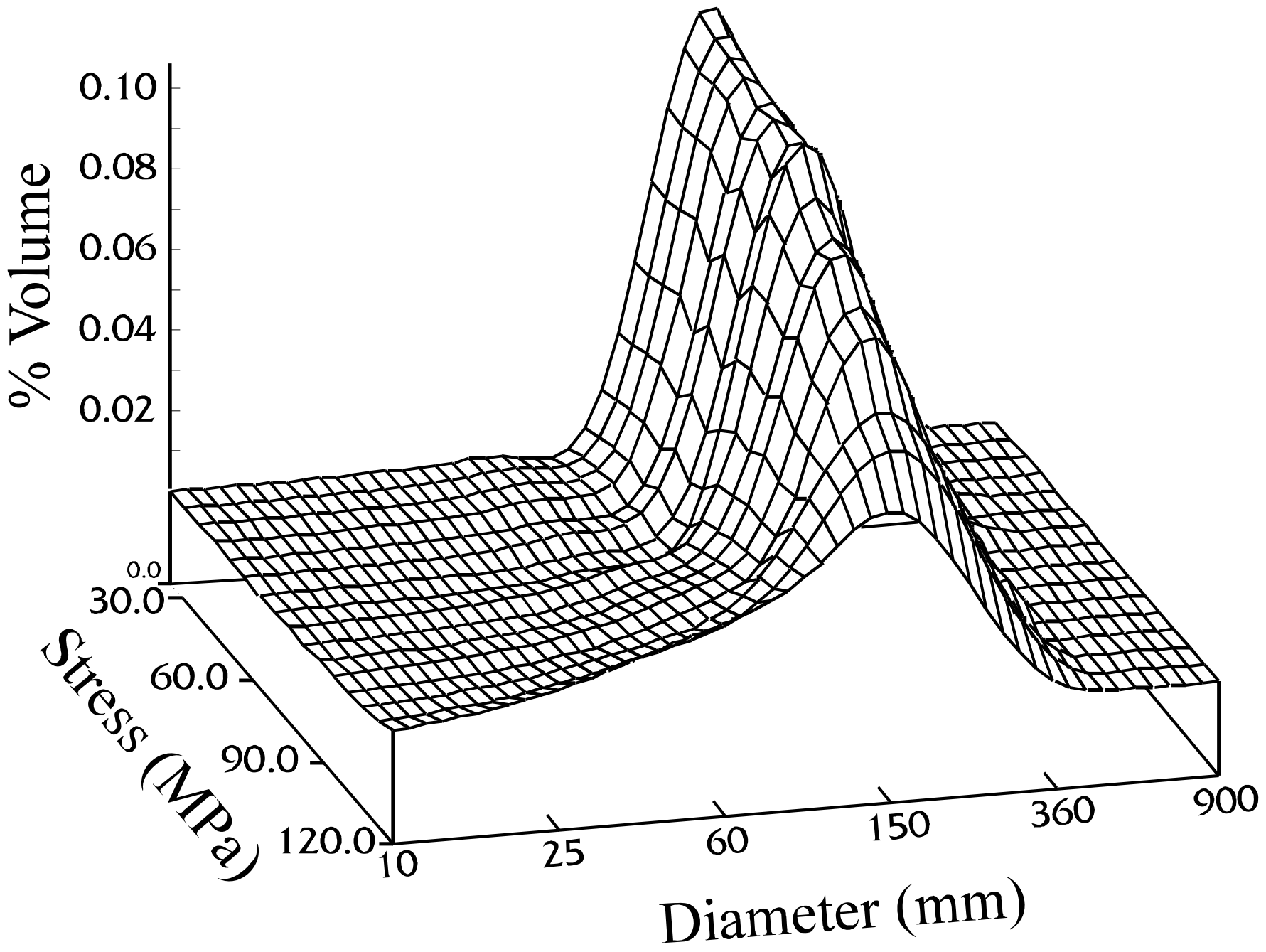

Figure 15 : *Grain-size distributions versus maximal axial stress (for all oedometric tests).*

Figure (16) gives the mean diameter of grains after testing for all the rigid confinement tests performed. The mean diameter for the virgin specimen is $d_{mean}$=196μm. After testing, the mean diameter may fall down to 65μm (values are given in [13]). For rigid confinement tests, a part of the specimen grains is crushed. The relationship between grains mean diameter after testing and axial stress is strong. It may point out that there is no dynamical effect in the comminution phenomena. This effect may be significant in the transient phase of the test, but it is not really possible to determine grain-size during testing.

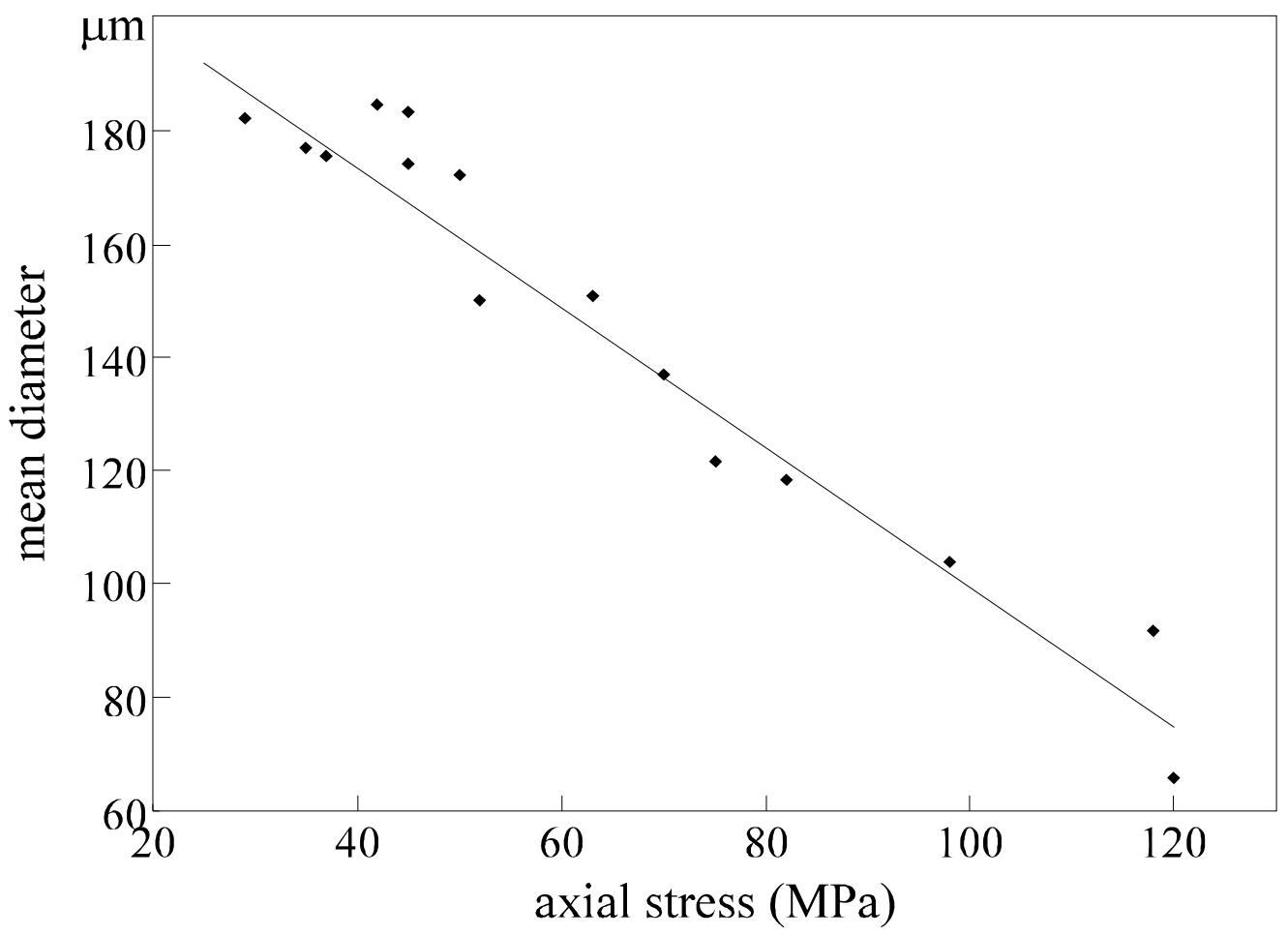

Figure 16 : *Mean diameter versus maximum axial stress (for all oedometric tests)*

## 6.2 Influence of fast loading

Shukla et al. [18] performed experiments on photoelastic grain packing to study the propagation of a loading wave in a granular medium. He evidences the influence on propagation of a hard grain or a void in 1D or 2D granular assemblies. For static loading on granular assemblies, many authors showed the appearance of force chains of different intensities [8].

In dynamic experiments, fast loading can generate "preferential force chains" of high intensity. Grains cannot choose a path of minimum force intensity to reach equilibrium. These "preferential force chains" give high stress intensity. It is interesting to determine what is the influence of this transient effect (strain rate effect) and what is the impact of axial stress for the whole test (stress level effect figure (16)).

### 6.3 Fracture energy

Considering the grain-size curves of figure (16), it is possible to make a quantitative analysis of the grains fracture in teh specimen.

Each grain-size curve may be related to the fracture energy of the corresponding test. Assuming spherical grains, we can write a simple relation between fracture energy $E_{fract}$ and new grain surface created. As indicated by Fukumoto [5,6], the expression of $E_{fract}$ takes one of the following form (proposed by Kick and Rittinger, see [5,6]) :

$$E_{Ritt} = C_R . \sum_i \left[ q_i \cdot \left( \frac{1}{x_i} - \frac{1}{x_0} \right) \right] \tag{7}$$

$$E_{Kick} = C_K . \sum_i \left[ q_i . \log \left( \frac{x_0}{x_i} \right) \right] \tag{8}$$

where $q_i$ is the part of grains of diameter $x_i$
and $x_o$ the mean grain diameter

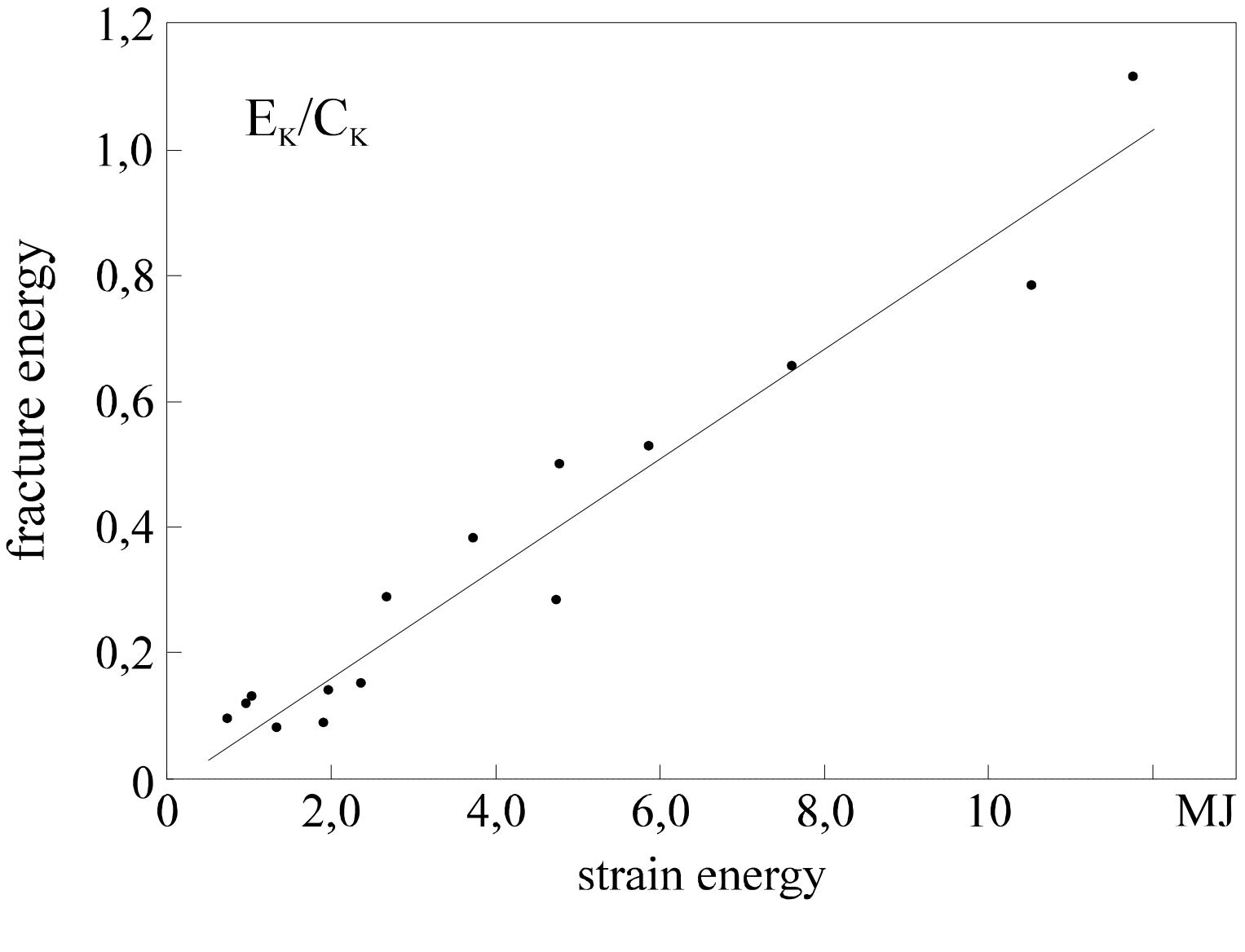

Figure 17 : *Fracture energy versus strain energy (for all oedometric tests)*

From (axial) stress-strain curves, *strain energy* dissipated in the specimen $E_{\sigma-\varepsilon}$ may be estimated. It is interesting to compare *fracture energy* and *strain energy*. As is is shown in figure (17), for all oedometric dynamic tests, *fracture energy is proportional to strain energy* $E_{\sigma-\varepsilon}$ . Thus there is a close relationship between *strain energy* dissipated in the specimen (estimated from the response) and *fracture energy* (computed from expression (8) and grain-size distributions). However, after loading of the specimen, a part of the stress waves is still travelling in the bars. Nevertheless, from this relationship between fracture energy and strain energy, it seems that the only first loading wave changes grain-size distributions. It is a logical conclusion considering velocities at both bar-specimen interfaces after unloading (see [13]). There is a separation between the bars and the specimen after the first unloading phase.

## 7. MAIN RESULTS

Using S.H.P.B loading to investigate dynamic response of soil appears to be a promising approach (see for example figure (3)). The most important result of this study is that it gives a mean to determine the whole 3D stress path (3D-S.H.P.B). An original experimental arrangement is proposed which we called : the 3D-Split Hopkinson Pressure Bar. The comparison of different tests using various confining conditions shows the strong influence of the loading path on the dynamic response. Calculation of dynamic pseudo Poisson's ratio reveals the strain rate effect on the 3D-dynamic response. Analysis at grain-size scale also gives interesting results about the relation between dynamic response and grain-size changes.

Further investigations could compare rigid (oedometric) and soft confining tests like ours with semi-rigid confining tests as Bragov [1] performed them (thin confining cylinder allowing (measured) radial strain). The real influence of strain rate on the dynamic response has to be accurately studied using different types of striker bar (length, mass...). The inverse problem approach has already given several interesting results about the transient phase [11] and is a promising tool for future research using S.H.P.B technique.

## 8. ACKNOWLEDGEMENTS

The authors thank F.Darve Head of *GRECO-Géomatériaux* and J.P.Touret Coordinator of the *Soil Dynamics Group* for their friendly support to this research.